\documentclass[a4paper,oneside]{article}

\usepackage{amsfonts,amssymb,amsthm}
\usepackage{latexsym}
\usepackage[mathscr]{eucal}

\def\mb#1{{\mathbf #1}}

\begin{document}

\title
{Self-dual Permutation Codes of Finite Groups
        in Semisimple Case
 \thanks{Supported by NSFC Grant No. 10871079.}
}
\author{Yun Fan, ~~ Guanghui Zhang
  \\[5pt]
 \small Department of Mathematics, Central China Normal University\\
  \small Wuhan, 430079, China\\
 \small Email: yunfan02@yahoo.com.cn\\[8pt]
 }

\date{}

\maketitle

\begin{abstract}
The existence and construction of self-dual codes in a permutation module
of a finite group for the semisimple case are described from two aspects,
one is from the point of view of
the composition factors which are self-dual modules,
the other one is from the point of view of the Galois group of the
coefficient field.

\end{abstract}

\section{Introduction}

Let $F$ be a finite field of order $q$ which is a power of a prime integer;
and let $X$ be a finite set. By $FX$ we denote the $F$-vector space
with the basis $X$ and with the usual scalar product
as its standard inner product.
Any subspace $C$ of $FX$ is just the usual linear code over $F$.
In coding-theoretic notation, with respect to the standard inner product,
the orthogonal subspace $C^\bot$ of a linear code $C$
is called the {\em dual code} of $C$; and
$C$ is called a {\em self-orthogonal code} if $C\subseteq C^\bot$;
and $C$ is called a {\em self-dual code} if $C=C^\bot$.

If $X$ is a group, then $FX$ is an algebra with multiplication
induced by the multiplication of the group $X$,
which is called the group algebra of the group $X$ over $F$;
and any left ideal $C$ of $FX$ is said to be a {\em group code}.
It is an interesting question to find conditions such that a group algebra
has a self-dual group codes. More generally, this question can be extended
to the group algebras over finite rings.

In \cite{R}, finite abelian groups are considered and some results on
the non-existence of self-dual group codes are shown.
For the direct product of a finite $2$-group and a finite $2'$-group,
reference \cite{Hug} showed when the self-dual group codes do not exist.
Using the representation theory of finite groups,
for group algebras over finite Galois rings
reference \cite{Wi} gave a complete answer for this question.
In particular, it is an easy conclusion that there is no self-dual
code for finite groups of odd order.

Thus it is reasonable to consider the self-dual
extended group codes for finite groups of odd order.
And \cite{MW} obtained some interesting conditions for the existence
of such self-dual codes in characteristic $2$:
one is from the point of view of self-dual modules,
another one is an elementary number-theoretical condition;
and \cite{MW} also showed some constructions of such codes.

Extending group codes, \cite{FY} discussed the so-called
{\em permutation codes} of finite groups.
If $G$ is a finite group and $X$ is a finite $G$-set, then
$FX$ is called a {\em permutation $FG$-module},
which has the standard inner product with respect to the basis $X$;
any $FG$-submodule $C$ of $FX$ is said to be an {\em $FG$-permutation code}.
If $X$ is a transitive $G$-set, the permutation cades of $FX$
is called {\em transitive permutation codes}.
View the base set of the group $G$ as a left regular $G$-set,
then the group codes are just the permutation codes of $FG$.
Some important codes are permutation codes in natural ways,
but may not be group codes; e.g. the so-called {\em multiple-cyclic codes};
see \cite{FY} for details. Moreover,
the research of permutation codes is of interests
from the point of view of automorphism groups of linear codes, for:
any permutation automorphism of a linear code is just a permutation
of the standard basis of the linear code.
In \cite{FY} some conditions are obtained for the non-existence of the self-dual
transitive permutation codes of finite groups.
And it is also an easy conclusion that there is no self-dual
transitive permutation code for finite groups of odd order.

In this paper we discuss the existence and construction of
self-dual permutation codes for the semisimple case.
The outline is as follows.

{\em Throughout the paper}, $F$ denotes a finite field of order $q$,
and $G$ denotes a finite group of order coprime to $q$,
and any $FG$-module is finite-dimensional.

In \S2, we first make observations on the related module-theoretical aspects,
and then turn to the permutation codes.
Since $FG$ is a semisimple algebra (Maschke's theorem),
any $FG$-module $V$ is decomposed into a direct sum
of irreducible $FG$-modules with
the collection of the irreducible summands
is unique determined up to isomorphism;
any irreducible $FG$-module $W$ which appears in the direct sum is called a
{\em composition factor} of $V$, and the number of the direct summands which
are isomorphic to $W$ is called the {\em multiplicity} of $W$ in $V$.
The dual space $V^*:={\rm Hom}_F(V,F)$ consisting of all the linear form
of $V$ is an $FG$-module with $G$-action:
$(g\varphi)(v)=\varphi(g^{-1}v)$, $\forall$
$g\in G$, $\varphi\in{\rm Hom}_F(V,F)$, $v\in V$.
We call $V$ a {\em self-dual $FG$-module} if $V\cong V^*$.
So, ``self-dual module'' and ``self-dual code'' are different concepts.
After the module-theoretical results which we need are obtained,
we turn to coding-theoretical notation, and show that,
for even $q$ and odd $|G|$, an $FG$-permutation module $FX$
has self-dual permutation codes if and only if
any self-dual composition factor of the $FG$-module $FX$
has even multiplicity.
For odd $q$, only a sufficient condition is obtained.

In \S3, we discuss transitive permutation codes, i.e.
codes of an permutation module $FX$ with a transitive $G$-set $X$.
We first reduce the existence of the so-called self-dual
{\em extended transitive permutation codes} to the
existence of such transitive permutation codes $C$ of $FX$
that $C^\bot=C\oplus F$.
And we show that, for a transitive $G$-set $X$ with length $n=|X|$,
if the integer $q$ as an element of the multiplicative group ${\Bbb Z}_n^\times$
has odd order, then there is a permutation code $C$ of $FX$ such that
$C^\bot=C\oplus F$.
It is easy to see that this elementary number-theoretical condition
is similar to that in \cite{MW}. However,
the situation of transitive permutation codes
is more delicate than that of group codes,
so that we take a way different from \cite{MW} to treat our cases;
and we obtained no necessary and sufficient conditions,
though some more results are shown in \S3 which seem interesting.

\section{Self-dual modules and self-dual codes}

We adopt the usual notation about linear forms, bilinear forms etc.
from the usual linear algebra.
A bilinear form $f(-,-)$ on an $FG$-module $V$
is said to be {\em $G$-invariant} if
$$
 f\big(g(u),\,g(v)\big)=f(u,\,v)\,,\qquad\forall~ u,v\in V\,.
$$

Let $V$ be an $FG$-module with a $G$-invariant
non-degenerate bilinear form $\langle-,-\rangle$.
Let $U$, $W$ be submodules of $V$. Denote
$$
 {\rm Ann}^l_W(U)=\{w\in W\mid\langle w,u\rangle=0,~\forall~u\in U\}\,,
$$
$$
 {\rm Ann}^r_W(U)=\{w\in W\mid\langle u,w\rangle=0,~\forall~u\in U\}\,;
$$
in particular, denote $U^\bot={\rm Ann}^r_V(U)$ and
${^\bot U}={\rm Ann}^l_V(U)$.
From the $G$-invariancy of $\langle-,-\rangle$, it is easy to see that
${\rm Ann}^l_W(U)$ and ${\rm Ann}^r_W(U)$ are $FG$-submodules.
Note that ${\rm Ann}^l_W(U)={\rm Ann}^r_W(U)$
and ${^\bot U}=U^\bot$ once $\langle-,-\rangle$ is symmetric.
For any $v_0\in V$ we have the linear form 
$\langle-, v_0\rangle: V\to F$, $v\mapsto \langle v,v_0\rangle$;
and restricting it to $U$, we have the linear form
$\langle -,v_0\rangle|_U$ on $U$ and it is easy to check that
\begin{equation}\label{V to U*}
 V\longrightarrow U^*,\quad v_0\longmapsto \langle -,v_0\rangle|_{U}
\end{equation}
is a surjective $FG$-homomorphism with kernel $U^\bot$;
thus we have an exact sequence of $FG$-homomorphisms:
\begin{equation}\label{UU* exact}
 0~\longrightarrow~U^\bot~\longrightarrow~V
    ~\longrightarrow~ U^* ~\longrightarrow~ 0\;;
\end{equation}
in particular, $\dim V=\dim U+\dim U^\bot$ because $\dim U=\dim U^*$.
Restricting the bilinear form $\langle-,-\rangle$ to the $FG$-submodule $U$,
we get a $G$-invariant symmetric bilinear form on $U$.
If the restricted bilinear form on $U$ is non-degenerate (equivalently,
${\rm Ann}^r_U(U)=U\cap U^\bot =0$),
we say that $U$ is a {\em non-degenerate submodule}.
On the other hand, if the restricted bilinear form on $U$ is zero
(equivalently, $U\subseteq U^\bot$),
we say, in module-theoretical notation,
that $U$ is an {\em isotropic submodule}.

\smallskip
Recall that any $FG$-module $V$ is written into a direct sum
of irreducible modules, and the irreducible direct summands
are partitioned by isomorphism, hence
$V=V_1\oplus\cdots\oplus V_h$, with every $V_i$ consisting of
the irreducible direct summands which are isomorphic
to one and the same irreducible module $W_i$,
but $V_i$ and $V_j$ for $i\ne j$ have no composition factors in common;
thus $V_i\cong m_iW_i$ with $m_i$ being the multiplicity of $W_i$ in $V$,
and $V_i$ is called the {\em homogeneous component} of $V$
associated with the irreducible module $W_i$,
and $V=V_1\oplus\cdots\oplus V_h$ is
called the {\em canonical decomposition} (or {\em homogeneous decomposition})
of $V$, see \cite[\S2.6]{S};
the canonical decomposition of $V$ is unique,
so that for any submodule $U$ of $V$ we have
\begin{equation}\label{canonical decomposition}
 U=(U\cap V_1)\oplus\cdots\oplus(U\cap V_h)\,.
\end{equation}

\smallskip\noindent{\bf Lemma 1.}\quad\it Let $V$ be an $FG$-module
with a $G$-invariant non-degenerate bilinear form;
and $U$ be an $FG$-submodule.

(1)\quad If $U$ is non-degenerate 
then $U$ is an self-dual $FG$-module.

(2)\quad If $U$ is irreducible,
then $U$ is either non-degenerate or isotropic.

(3)\quad If $U$ is a homogeneous component associated
with an irreducible module $W$, then
$W$ is self-dual if and only if $U$ is non-degenerate.
$W$ is not self-dual if and only if $U$ is isotropic.
\rm

\smallskip{\bf Proof.}\quad
(1).\quad The non-degeneracy of $U$ implies
$U\cap U^\bot=0$; thus from that $\dim V=\dim U+\dim U^\bot$ we get
$V=U^\bot\oplus U$, and it follows from the exact sequence (\ref{UU* exact})
that $U\cong V/U^\bot\cong U^*$.

(2).\quad Because $U\cap U^\bot$ is an $FG$-submodule of $U$,
the irreducibility of $U$ implies that
either $U\cap U^\bot=0$ or $U\cap U^\bot=U$.

(3).\quad From the exact sequence (\ref{UU* exact}) and the semi-simplicity,
we have that $V=U^\bot\oplus U'$ with $U'\cong U^*$.
Since $FG$ is an Frobenius algebra,
it is known (e.g. see \cite{Wo})
that the dual modules of all the composition factors of $U$ are just
all the composition factors of $U^*$. Thus $U'$ is a homogeneous component too.
Thus the conclusions follows from the uniqueness of the
homogeneous decomposition.

\smallskip{\bf Remark.}\quad
It is well-known that
``there is a $G$-invariant non-degenerate bilinear form
on a $FG$-module $V$ if and only if $V$ is a self-dual $FG$-module''.
The necessity is a special case of Lemma 1(1);
and the sufficiency follows that, with an $FG$-isomorphism $\alpha: V\to V^*$,
the composition map
$$\begin{array}{ccccc}
 V\times V &\longrightarrow& V^*\times V &\longrightarrow & F\,,\\
 (v,v') &\longmapsto& (\alpha(v),\;v') & \longmapsto& \alpha(v)(v')\,.
 \end{array}
$$
is a $G$-invariant non-degenerate bilinear form on $V$.
For more details, please see \cite[Ch.VII, \S8]{HB}.

\smallskip\noindent{\bf Lemma 2.}\quad\it Let
$V$ be an $FG$-module with a $G$-invariant non-degenerate
symmetric bilinear form;
let $U$ be an isotropic $FG$-submodule of $V$.
Then the following are equivalent:

(i)\quad $U^\bot=U$;

(ii)\quad $\dim U = \dim V/2$;

(iii)\quad the collection of the composition factors of $U$
and the dual modules of the composition factors of $U$
is the collection of the composition factors of $V$.
\smallskip\rm

{\bf Proof.}\quad
(i) $\Leftrightarrow$ (ii) is obvious
since $\dim V= \dim U^\bot+\dim U$.

(i) $\Leftrightarrow$ (iii). Similar to the proof for Lemma 1(3),
$V=U^\bot\oplus U'$ with $U'\cong U^*$;
but now $U\subseteq U^\bot$ by hypothesis,
so the equivalence is obvious.

Recall from the usual linear algebra that, for an $FG$-module $V$,
any bilinear form $f$ on $V$ corresponds to exactly one
linear form $\bar f$ on the tensor product space $V\otimes_F V$:
$\bar f(v\otimes v')=f(v,v')$; in other words, the dual space
$(V\otimes_F V)^*$ is identified
with the space of all the bilinear forms on $V$.
As usually, $V\otimes_F V$ is an $FG$-module by diagonal action of $G$,
hence $(V\otimes_F V)^*$ is also an $FG$-module by diagonal action of $G$;
and the space of all the $G$-invariant bilinear forms
is identified with the subspace of all the $G$-fixed points
of $(V\otimes_F V)^*$, denoted by $((V\otimes_F V)^*)^G$.

On the other hand, $G$ acts on the space ${\rm Hom}_F(V,V)$ of all the
linear transformations of $V$ in the following way:
$$(g\alpha)(v)=g(\alpha(g^{-1}v)\,,\qquad
  \forall~ g\in G,~ \alpha\in{\rm Hom}(V,V),~v\in V\,;$$
and the subspace ${\rm Hom}_{FG}(V,V)$ of all the $FG$-endomorphisms of $V$
is just the set of all the $G$-fixed points of ${\rm Hom}_F(V,V)$.

\smallskip\noindent{\bf Lemma 3.}\quad\it
Let $V$ be an $FG$-module with a $G$-invariant
non-degenerate symmetric bilinear form $\langle-,-\rangle$.
For any linear transformation $\alpha\in{\rm Hom}_F(V,V)$ define
$$
 \varphi_\alpha(u,\, v)=\langle\alpha(u),\, v\rangle\,,\qquad\forall~
  u,v\in V\,.
$$
Then $\varphi_\alpha$ is a bilinear form on $V$, and
$$
 \varphi:\quad {\rm Hom}_F(V,V)~\longrightarrow~ (V\otimes_F V)^*,\quad
  \alpha\longmapsto\varphi_\alpha\,.
$$
is an $FG$-isomorphism, and:

(1)\quad $\varphi_\alpha$ is $G$-invariant
if and only if $\alpha$ is an $FG$-endomorphism;

(2)\quad $\varphi_\alpha$ is non-degenerate
if and only if $\alpha$ is a non-degenerate transformation;

(3)\quad $\varphi_\alpha$ is a symmetric
if and only if $\alpha$ is a symmetric transformation.\smallskip\rm

{\bf Proof.}\quad
It is easy to check that $\varphi_\alpha$ is a bilinear form on $V$,
and that $\varphi$ is a linear map;
and that $\varphi$ is injective because $\langle-,-\rangle$
is non-degenerate, hence $\varphi$ is bijective since
$\dim {\rm Hom}_F(V,V)=\dim(V\otimes_F V)^*$.
Next, for any $g\in G$, any $\alpha\in{\rm Hom}_F(V,V)$,
and any $u,v\in V$, we have
\begin{eqnarray*}
 \varphi_{g\alpha}(u\otimes v)&=&\langle(g\alpha)(u),\,v\rangle
 =\langle g\alpha(g^{-1}u),\,v\rangle=\langle \alpha(g^{-1}u),\,g^{-1}v\rangle\\
 &=&\varphi_\alpha(g^{-1}u\otimes g^{-1}v)=\varphi_\alpha(g^{-1}(u\otimes v))
 =(g\varphi_\alpha)(u\otimes v)\,.
\end{eqnarray*}
So $\varphi$ is an $FG$-isomorphism. Hence we have the following isomorphism
\begin{equation}\label{Hom-Tensor}
 {\rm Hom}_{FG}(V,V)~\buildrel\cong\over\longrightarrow~((V\otimes_F V)^*)^G,
 \quad \alpha~\longmapsto~\varphi_\alpha\,;
\end{equation}
that is, (1) holds. The (2) and (3) can be verified straightforwardly.
\smallskip

Let $V$ and $V'$ be $FG$-modules equipped with $G$-invariant bilinear forms
$f$ and $f'$ respectively. We say that an $FG$-homomorphism
$\alpha: V\to V'$ is compatible with the bilinear forms $f$ and $f'$ if
$f'(\alpha(u),\,\alpha(v))=f(u,\,v)$ for all $u,v\in V$.

If $f$ is a non-degenerate bilinear form on $V$, then
any $FG$-homomorphism $\alpha: V\to V'$
which is compatible with $f$ and $f'$ must be injective; for:
$\alpha(u)=0$ implies that for any $v\in V$ we have
that $f(u,\,v)=f'(\alpha(u),\,\alpha(v))=f'(0,\,\alpha(v))=0$,
hence $u=0$ by the non-degeneracy of the form $f$.

\smallskip\noindent{\bf Lemma 4.}\quad\it
Assume that $q$ is even, and $V$ is a
self-dual irreducible $FG$-module.
If both $f$ and $f'$ are $G$-invariant non-degenerate symmetric bilinear
forms on $V$, then there is an $FG$-automorphism
$\beta: V\to V$ which is compatible with $f$ and $f'$. \smallskip\rm

{\bf Proof.}\quad
Apply the isomorphism (\ref{Hom-Tensor})
to the $FG$-module $V$ with the $G$-invariant
non-degenerate symmetric bilinear form $f$.
Since $V$ is irreducible, by the Schur's lemma,
$\tilde F:={\rm Hom}_{FG}(V,V)$ is a finite dimensional division $F$-algebra,
hence $\tilde F$ is a field extension of $F$ as it is finite.
By the commutativity of $\tilde F$, it is easy to check that
the sum and the product of any two symmetric transformations in $\tilde F$
are still symmetric transformations, so
all the symmetric transformations in $\tilde F$ form a
subfield $\hat F$ of $\tilde F$.

By Lemma 3, for the $G$-invariant non-degenerate symmetric
bilinear form $f'$, there is an $\alpha\in\hat F-\{0\}$ such that
$$ f'(u,v)=\varphi_\alpha(u,v)=f\big(\alpha(u),v\big)\,,\qquad
 \forall~ u,v\in V\,.$$
Since $\hat F$ is a finite field of characteristic $2$,
the map $\hat F\to\hat F$, $\lambda\mapsto\lambda^2$,
is an automorphism of $\hat F$.
So there is a $\beta\in\hat F$ such that $\beta^2=\alpha^{-1}$.
Then $\beta:V\to V$ is an $FG$-automorphism of $V$ and
a symmetric transformation with respect to the bilinear form $f$;
and, noting that $\alpha\beta=\beta\alpha$,
for any $u,v\in V$ we have
$$
 f'\big(\beta(u),\,\beta(v)\big)=f\big(\alpha(\beta(u)),\,\beta(v)\big)
 =f\big((\beta\alpha\beta)(u)),\,v\big)=f\big(u,\,v\big)\,.
$$
That is, $\beta$ is compatible with the bilinear form $f$ and $f'$.

\smallskip\noindent{\bf Theorem 1.}\quad\it
Let $F$ be a finite field of characteristic $2$
 and $G$ be a finite group of odd order.
Let $V$ be an $FG$-module with
a $G$-invariant non-degenerate symmetric bilinear form.
Then the following are equivalent:

(i)\quad every self-dual composition factor of $V$
has even multiplicity;

(ii)\quad there is an $FG$-submodule $U$ of $V$ such that $U^\bot=U$.
\smallskip\rm

{\bf Proof.}\quad We denote $\langle-,-\rangle$ for the
$G$-invariant non-degenerate symmetric bilinear form on $V$.

(ii) $\Rightarrow$ (i).\quad
This is an easy consequence of Lemma 2 (i)$\Rightarrow$(iii).

(i) $\Rightarrow$ (ii).\quad 
Let $W$ be an irreducible $FG$-submodule of $V$.

Case 1: $W\subseteq W^\bot$.
By the exact sequence (\ref{UU* exact}),
we have a submodule $W'$ of $V$ such that
$V=W^\bot\oplus W'$ and the homomorphism (\ref{V to U*}) induces
an isomorphism
$$ W'~\buildrel\cong\over\longrightarrow~ W^*,\quad
  w'\longmapsto\langle w',-\rangle|_{W}\,. $$
Therefore, the matrix of the symmetric bilinear form
$\langle-,-\rangle|_{W'\oplus W}$ restricted to $W'\oplus W$
is as follows
$$
 \pmatrix{0 & A\cr A^T &*}
$$
where $A$ is the matrix of the bilinear form
$W'\times W\to F$, $(w',w)\mapsto \langle w',w\rangle$
and $A^T$ denotes the transpose of $A$;
so $A$ is invertible, and
hence $W'\oplus W$ is a non-degenerate submodule of $V$.
Then
$$ V=(W'\oplus W)\oplus(W'\oplus W)^\bot$$
and $(W'\oplus W)^\bot$ is also non-degenerate submodule.

If $W$ is not a self-dual module, then $W'\cong W^*$ is not self-dual,
and hence $(W'\oplus W)^\bot$ also satisfies the condition (i).
Otherwise, $W$ is a self-dual module,
and $W'\cong W^*\cong W$ is a self-dual module too,
hence $(W'\oplus W)^\bot$ still satisfies the condition (i).
In a word, by induction, there is a submodule $S$ of
$(W'\oplus W)^\bot$ such that ${\rm Ann}_{(W'\oplus W)^\bot}(S)=S$.
Take $U=W\oplus S$; then it is easy to check
that $U^\bot =U$ and (ii) holds.

\smallskip
Case 2: $W\not\subseteq W^\bot$.
Then $W$ is non-degenerate, i.e. $V=W\oplus W^\bot$,
and $W$ is a self-dual module, see Lemma 1(2).
By the condition (i),
there is a direct decomposition $W^\bot=\tilde W\oplus U$
such that $\tilde W\cong W$, and $V=W\oplus \tilde W\oplus U$.

If $\tilde W\subseteq \tilde W^\bot$,
then it is reduced to Case 1 and the (ii) holds by induction.
So we assume that $\tilde W\not\subseteq \tilde W^\bot$,
and hence $\tilde W$ is also non-degenerate.
Since $W\bot\tilde W$,
the submodule $W\oplus\tilde W$ is non-degenerate too.

Let $f$ and $\tilde f$ denote the restrictions of $\langle-,-\rangle$
on $W$ and on $\tilde W$ respectively; so $f$ and $\tilde f$
are $G$-invariant non-degenerate symmetric bilinear forms on
$W$ and $\tilde W$ respectively.
Let $\alpha: W\to\tilde W$ be an $FG$-isomorphism.
Then $\alpha$ induces a $G$-invariant non-degenerate symmetric bilinear
form $f'$ on $W$ as follows:
$$ f'(u,w):=\tilde f\big(\alpha(u),\,\alpha(w)\big)\,,\qquad
    \forall~u,w\in W\,.
$$
By Lemma 4, there is an $FG$-automorphism
$\beta: W\to W$ which is compatible with $f$ and $f'$, i.e.
$$
 f'\big(\beta(u),\,\beta(w)\big)=f(u,\,w)\,,\qquad
  \forall~ u,w\in W\,.
$$
Let $\gamma=\alpha\beta$.
Then $\gamma: W\to\tilde W$ is an $FG$-isomorphism, and
for any $u,w\in W$ we have
$$
 \tilde f\big(\gamma(u),\,\gamma(w)\big)
 =\tilde f\big(\alpha(\beta(u)),\,\alpha(\beta(w))\big)
 =f'\big(\beta(u),\,\beta(w)\big)=f(u,\,w)\,;
$$
that is, $\gamma$ is an $FG$-isomorphism compatible with
the bilinear forms $f$ and $\tilde f$. Let
$$ W'=\{w+\gamma(w)\mid w\in W\}\subseteq W\oplus \tilde W\,. $$
It is a routine to check that $W'$ is a submodule and $W'\cong W$;
but, noting that $W\bot \tilde W$ and ${\rm char}\,F=2$,
for any $u+\gamma(u)\in W'$ and $w+\gamma(w)\in W'$ with $u,w\in W$ we have
\begin{eqnarray*}
\big\langle u+\gamma(u),\;w+\gamma(w)\big\rangle &=&
\big\langle u,\,w\big\rangle +\big\langle\gamma(u),\,\gamma(w)\big\rangle\\
&=&f(u,\,w)+\tilde f\big(\gamma(u),\,\gamma(w)\big)\\
&=&f(u,\,w)+f(u,\,w)=0\,.
\end{eqnarray*}
So $W'\cong W$ is an irreducible $FG$-submodule of $V$
and $W'\subseteq W'^\bot$, and it is reduced to the Case 1 and
(ii) holds by induction again.

\smallskip{\bf Remark.}\quad In the proof of Theorem 1,
Lemma 4 is quoted only in Case 2 where $W$ and $\tilde W$ are
self-dual composition factors of $V$. Thus, as a consequence of the proof,
we have the following conclusion.

\smallskip\noindent{\bf Proposition 1.}\quad\it
Let $G$ be a finite group of order coprime to the
characteristic (not necessary $2$) of the finite field $F$,
and $V$ be an $FG$-module with a $G$-invariant non-degenerate
symmetric bilinear form. If $V$ has no self-dual composition factor,
then $V$ has a submodule $U$ such that $U^\bot =U$.\rm

\smallskip
Now we turn to permutation codes.
Let $X$ be a finite set;
by ${\rm Sym}(X)$ we denote the group of all the permutations of $X$.
If there is a group homomorphism $G\to{\rm Sym}(X)$,
then $X$ is called a {\em $G$-set}. In that case,
any $g\in G$ is mapped to a permutation: $X\to X$, $x\mapsto gx$.
Hence, $(gg')x=g(g'x)$ for all $g,g'\in G$ and $x\in X$;
and $1x=x$ for all $x\in X$.

Let $FX=\{\,\sum_{x\in X} a_x\,x\;|\;a_x\in F\;\}$
be the vector space over $F$ with basis $X$.
Extending the $G$-action on $X$ linearly, $FX$ becomes
an $FG$-module, called an {\em $FG$-permutation module}
with permutation basis $X$, please cf. \cite[\S12]{A}.

We say that $C$ is an {\em $FG$-permutation code} of $FX$,
denoted by $C\le FX$, if $C$ is an $FG$-submodule
of the $FG$-permutation module $ FX$; and
a permutation code $C$ is said to be {\em irreducible}
if $C$ is an irreducible $FG$-submodule of $FX$.
Further, if $X$ is a transitive $G$-set, then any
$FG$-permutation code $C$ of $FX$ is said to be a
{\em transitive permutation code}.

Recall that, for a linear code $C$ of length $n$ over $F$,
a permutation of the components of a word of length $n$ is said to be a
{\em permutation automorphism} of $C$ if the permutation keeps every code word
of $C$ still a code word. By ${\rm PAut}(C)$ we
denote the automorphism group of $C$ consisting of all
the permutation automorphisms of $C$.
It is easy to see that $C$ is an $FG$-permutation code
of a $G$-permutation set $X$ of cardinality $n$
if and only if there is a group homomorphism of $G$ to
${\rm PAut}(C)$.

There is a so-called scalar product of any two words of $FX$ as follows:
$$
 \left\langle \mb w,\;\mb w' \right\rangle
  = \sum_{x\in X}w_xw'_x\;,
 \quad \forall~ \mb w=\sum_{x\in X} w_x x,\;
                 \mb w'=\sum_{x\in X} w'_x x\in FX\;,
$$
which is obvious a non-degenerate symmetric bilinear form on $FX$,
we call it the {\em standard inner product} on $FX$
with respect to the permutation basis $X$.
Moreover, the standard inner product is $G$-invariant, since
for any $g\in G$ and any words $\mb w=\sum_{x\in X} w_x x$ and
 $\mb w'=\sum_{x\in X} w'_x x$ of $FX$, we have
\begin{eqnarray*}
\langle g(\mb w),\,g(\mb w')\rangle&=&
\left\langle g\Big(\sum_{x\in X} w_x x\Big),\,
   g\Big(\sum_{x\in X} w'_x x\Big)\right\rangle \\
 &=& \left\langle \sum_{x\in X} w_x(gx),\,
      \sum_{x\in X} w'_x(gx)\right\rangle = \sum_{x\in X} w_xw'_x\\
 &=& \langle\mb w,\,\mb w'\rangle\;;
\end{eqnarray*}
equivalently,
$$
 \langle g(\mb w),\,\mb w'\rangle=\langle\mb w,\,g^{-1}(\mb w')\rangle\;,\qquad
  \forall\;g\in G\,,\;\forall\;\mb w,\mb w'\in FX\;.
$$

Thus, $FX$ is a self-dual $FG$-module.
In fact, we can make the duality more precisely.
Just like the formula (\ref{V to U*}),
the standard inner product induces an isomorphism
$$
 FX~\buildrel\cong\over\longrightarrow~(FX)^*\,,\quad
   \mb u~\longmapsto~ \mb u^*:=\langle\mb u,-\rangle~,
$$
where
$$
 \mb u^*:\quad FX\longrightarrow F\,,\quad
  \mb w\longmapsto \mb u^*(\mb w)=\langle\mb u,\mb w\rangle~;
$$
and
$$
 X^*:=\{x^*\mid x\in X\}
$$
is a $G$-set with $G$-action
$$
 g(x^*)=(g^{-1}x)^*\,,\qquad\forall~ g\in G\,,~ x^*\in X^*~,
$$
such that $(FX)^*$ is an $FG$-permutation module of the $G$-set $X^*$,
and $\mb u\mapsto \mb u^*$ is a permutation isomorphism.

\smallskip Let $FX$ be an $FG$-permutation module.
For any permutation code $C$ of $FX$, since $C$ is an $FG$-submodule,
$C^\bot=\{\mb w\in FX\mid \langle\mb c,\,\mb w\rangle=0\,,
  \;\forall\; \mb c\in C\}$
is an $FG$-submodule again, i.e. $C^\bot$ is a permutation code again.
In coding-theoretical notation,
$C^\bot$ is said to be the {\em dual permutation code} of $C$.

An $FG$-permutation code $C\le  FX$ is said to be
{\em self-orthogonal} if $C\subseteq C^\bot$.
And a permutation code $C\le FX$ is said to be
{\em self-dual} if $C=C^\bot$.

With the coding-theoretical notation introduced above,
from Theorem 1 and Proposition 1, we have the following results at once.

\smallskip\noindent{\bf Theorem 2.}\quad\it
Let $F$ be a finite field of characteristic $2$,
and $G$ be a finite group of odd order, and $X$ be a finite $G$-set.
Then the following are equivalent:

(i)\quad every self-dual composition factor of $FX$ has even multiplicity;

(ii)\quad there is a self-dual $FG$-permutation code $C$ of $FX$.
\smallskip\rm

\smallskip\noindent{\bf Proposition 2.}\quad\it
Let $G$ be a finite group of order coprime to the
characteristic (not necessary $2$) of the field $F$,
and $X$ be a finite $G$-set.
If $FX$ has no self-dual composition factor,
then there is a self-dual $FG$-permutation code of $FX$.\rm

\section{Self-dual extended\\ transitive permutation codes}

If a $G$-set $X=\{x_0\}$ contains of only one element,
then $X$ is said to be the trivial $G$-set
and the permutation module $FX\cong F$ is just the {\em trivial $FG$-module},
which is obviously a self-dual module.

An elementary known fact is that, in the semisimple case,
for any transitive $G$-set $X$ the trivial $FG$-module $F$
is a composition factor of multiplicity $1$
of the $FG$-permutation module $FX$;
e.g. see \cite[Lemma 1]{FY}; we denote the
unique trivial submodule of $FX$ by $F$ if there is no confusion,
thus $FX=F\oplus F^\bot$. By Theorem 1, $FX$ has no self-dual codes.

Let $X$ be a transitive $G$-set. Let $\hat X=X\bigcup\{x_0\}$ be the
disjoint union of $X$ with a trivial $G$-set $\{x_0\}$, i.e. $x_0\notin X$.
Then $F\hat X=FX\oplus Fx_0$,
and any permutation code $C$ of $F\hat X$ is said to be
an {\em extended transitive permutation code} of $FX$.

\smallskip\noindent{\bf Lemma 5.}\quad\it Notation as above, and let
$n=|X|$. The following are equivalent:

(i)\quad there is a permutation code $C$ of $FX$ such that
$C^\bot=C\oplus F$ and, as an element of the field $F$,
 $-n$ has a square root in $F$;

(ii)\quad there is a self-dual permutation code $\hat C$ of $F\hat X$.
\rm

{\bf Proof.}\quad Let $e=\sum_{x\in X}x$;
then $Fe$ is the unique submodule of $FX$ which is isomorphic to $F$,
so $Fx_0\oplus Fe$ is the homogeneous component of $F\hat X$ associated with
the trivial module $F$.
Noting that $Fx_0\bot Fe$ and $\langle x_0,x_0\rangle =1$ and
$\langle e,\,e\rangle =n\ne 0$
(because $n\,\big|\,|G|$ which is coprime to $q=|F|$),
we see that $Fx_0\oplus Fe$ is a non-degenerate submodule of $F\hat X$.
Thus
$$ F\hat X = (Fx_0\oplus Fe)\oplus (Fx_0\oplus Fe)^\bot $$
and
$$(Fx_0\oplus Fe)^\bot=(Fx_0)^\bot\cap(Fe)^\bot=FX\cap(Fe)^\bot
  ={\rm Ann}_{FX}(Fe)\,.$$

(ii) $\Rightarrow$ (i).\quad
By the formula (\ref{canonical decomposition}) we have
$$
 \hat C=\big(\hat C\cap(Fx_0\oplus Fe)\big)
  \oplus\big(\hat C\cap {\rm Ann}_{FX}(Fe)\big).
$$
From the condition (ii) that $\hat C^\bot =\hat C$, by Lemma 2(ii), we have
$$
 \dim\big(\hat C\cap(Fx_0\oplus Fe)\big)=1\,,\qquad
  \dim\big(\hat C\cap {\rm Ann}_{FX}(Fe)\big)=\frac{n-1}{2}\,.
$$
Set $C=\hat C\cap {\rm Ann}_{FX}(Fe)$; it is easy to check
that, $C$ is a permutation code of $FX$ and $C^\bot=C\oplus Fe$ in $FX$.
On the other hand, for $C\cap(Fx_0\oplus Fe)$ which is a one-dimensional subspace,
we assume that $\lambda\in F$ such that
$$
 \hat C\cap(Fx_0\oplus Fe)=F\cdot(\lambda x_0+e)\,;
$$
then $\langle \lambda x_0+e,\;\lambda x_0+e\rangle=0$;
i.e.
$$
 0=\langle \lambda x_0,\;\lambda x_0\rangle
   +\langle e,\,e\rangle=\lambda^2+n\,;
$$
that is, $\lambda^2=-n$.

(i) $\Rightarrow$ (ii).\quad In $FX$, since $\dim C+\dim C^\bot =n$,
from the condition that $C^\bot=C\oplus Fe$
we have that $\dim C=\frac{n-1}{2}$.
Turn to $F\hat X$, set $\lambda\in F$ such that $\lambda^2=-n$
and $\hat C:=F\cdot(\lambda x_0+e)\oplus C$;
as shown above, the $1$-dimensional submodule $F\cdot(\lambda x_0+e)$
of $Fx_0\oplus Fe$ is isotropic,
hence $\hat C$ is an isotropic submodule.
But $\dim\hat C=\frac{n+1}{2}$; and by Lemma 2,
$\hat C$ is a self-dual permutation code of $F\hat X$.

\smallskip{\bf Remark.}\quad 
In the above lemma, the condition ``$-n$ has a square root in $F$'' in (i)
always satisfies for characteristic $2$.

\smallskip
For any positive integer $n$ we denote ${\Bbb Z}_n$ the
residue ring of the integer ring ${\Bbb Z}$ modulo $n$, and denote
${\Bbb Z}_n^\times$ the multiplicity group consisting of all
the invertible elements of ${\Bbb Z}_n$.
So $q$ is considered as an element of ${\Bbb Z}_n^\times$,
and we can speak of the order of $q$ in the group ${\Bbb Z}_n^\times$.

\smallskip\noindent{\bf Lemma 6.}\quad\it Let $n$ be an odd
integer coprime to $q$. The following are equivalent:

(i)\quad The order of $q$ in ${\Bbb Z}_n^\times$ is odd.

(ii)\quad For any prime $p|n$
the order of $q$ in ${\Bbb Z}_p^\times$ is odd.\rm

\smallskip{\bf Proof.}\quad Let $n=p_1^{m_1}\cdots p_k^{m_k}$.
By Chinese Remainder Theorem we have the following
isomorphism about the multiplicative groups:
$$
 {\Bbb Z}_n^\times~\buildrel\cong\over\longrightarrow~
   {\Bbb Z}_{p_1^{m_1}}^\times\times\cdots\times
      {\Bbb Z}_{p_k^{m_k}}^\times\,,\quad
  a \longmapsto (a,\cdots,a)
$$
The order of $q\in {\Bbb Z}_n^\times$ is odd if and only if
the order $q\in {\Bbb Z}_{p_i^{m_i}}^\times$ is odd for every
$i=1,\cdots,k$. Consider the exact sequence of multiplication groups:
$$
 1~\longrightarrow~ 1+p_i{\Bbb Z}_{p_i^{m_i}}
 ~\buildrel{\rm incl}\over\longrightarrow~{\Bbb Z}_{p_i^{m_i}}^\times
 ~\buildrel{\rho}\over\longrightarrow~{\Bbb Z}_{p_i}^\times
 ~\longrightarrow~ 1
$$
where ``incl'' is the inclusion map and $\rho$ is the natural map:
$$
 {\Bbb Z}_{p_i^{m_i}}^\times~\longrightarrow~{\Bbb Z}_{p_i}^\times\,,
 \quad a\longmapsto a\,.
$$
Since the order $|1+p_i{\Bbb Z}_{p_i^{m_i}}|=p_i^{m_i-1}$ is odd,
the order of $q\in{\Bbb Z}_{p_i^{m_i}}^\times$ is odd if and only if
the order of $q\in{\Bbb Z}_{p_i}^\times$ is odd.

\smallskip
Recall that $F$ is a finite field of order $q$.
For any positive integer $n$, in a suitable extension we can take a
primitive $n$'th root $\xi_n$ of unity,
and the extension $F(\xi_n)$ is independent of the choice of $\xi_n$;
and the order of the Galois group
$\left|{\rm Gal}\big(F(\xi_n)/F\big)\right|=|F(\xi_n):F|$
is just the order of $q$ in the multiplicative group ${\Bbb Z}_n^\times$.
As a consequence we have the following at once.

\smallskip\noindent{\bf Corollary 1.}\quad\it Let $n$ be an odd
integer coprime to $q$. The following are equivalent:

(i).\quad The extension degree $|F(\xi_n):F|$ is odd.

(ii).\quad For any prime $p|n$
the extension degree $|F(\xi_p):F|$ is odd.\rm

\smallskip
Let $H$ be a subgroup of the group $G$, and let $Y$ be a finite $H$-set;
then $FY$ is an $FH$-permutation module. We have the induced $FG$-module
$$ {\rm Ind}_H^G(FY)=FG\bigotimes_{FH}FY=\bigoplus_{t\in T} t\otimes FY~,$$
where $T$ is a representative set of the left cosets of $G$ over $H$;
and ${\rm Ind}_H^G(FY)$ is a vector space with basis
$$ X:={\rm Ind}_H^G(Y)=\bigcup_{t\in T}t\otimes Y
 =\bigcup_{t\in T}\left\{t\otimes y\mid y\in Y\right\} $$
which is a $G$-set with $G$-action as follows:
$$ g(t\otimes y)= t_g\otimes t_g^{-1}gty\,,
  \qquad\forall~\; g\in G,~ t\in T,~ y\in Y\,, $$
where $t_g$ is the representative of the unique left coset $t_gH$ such that
$gt\in t_gH$, or equivalently $t_g^{-1}gt\in H$.
We say that ${\rm Ind}_H^G(FY)$ is the {\em induced $FG$-permutation module}
with the {\em induced $G$-set} ${\rm Ind}_H^G(Y)$.

\smallskip\noindent{\bf Lemma 7.}\quad\it
Notation as above; and let $D$ be an $FH$-permutation code of the
$FH$-permutation module $FY$; then
$$
 {\rm Ind}_H^G(D)^\bot={\rm Ind}_H^G(D^\bot)\,.
$$\rm

{\bf Proof.}\quad It is obvious that the induced module
$C:={\rm Ind}_H^G(D)$ is a submodule of
${\rm Ind}_H^G(FY)=\bigoplus_{t\in T}t\otimes FY$,
and we have a direct decomposition of $F$-spaces:
$$ {\rm Ind}_H^G(D)=\bigoplus_{t\in T}t\otimes D~, $$
where each $t\otimes D$ is an $F$-subspace of $t\otimes FY$.
Each $t\otimes FY$ is an $F$-space with bases $t\otimes Y$,
hence with the standard inner product:
$$
\Big\langle\sum_{y\in Y}a_y(t\otimes y),~\sum_{y\in Y}b_y(t\otimes y)\Big\rangle
 =\sum_{y\in Y} a_yb_y\,;
$$
and
$$
 FY~\longrightarrow~ t\otimes FY\,,\quad
 \sum_{y\in Y}a_yy~\longmapsto~\sum_{y\in Y}a_y(t\otimes y)\,,
$$
is an isometric $F$-isomorphism. With respect to the isometries,
it is clear that $(t\otimes D)^\bot=t\otimes D^\bot$; hence
$$
 {\rm Ind}_H^G(D)^\bot=\bigoplus_{t\in T} (t\otimes D)^\bot
  =\bigoplus_{t\in T}t\otimes D^\bot={\rm Ind}_H^G(D^\bot)~.
$$

\noindent{\bf Lemma 8.}\quad\it
Let $p$ be an odd prime which is coprime to $q$ such that
the order of $q$ in ${\Bbb Z}_p^\times$ is odd.
Let $A$ be a finite abelian $p$-group, and
$H$ be a finite group of odd order which acts on the group $A$.
Then there is a group code $C\le FA$ which is stable by the action of $H$
and $C^\bot=C\oplus F$,
where $F$ denotes the unique trivial module of $FA$.\rm

\smallskip{\bf Proof.}\quad Let $|A|=n$ which is a power of $p$;
take a primitive $n$'th root $\xi$ of unity, and denote $\tilde F=F(\xi)$.
Then $\tilde F A$ is a splitting semisimple commutative algebra.
Let $\Gamma={\rm Gal}(\tilde F/F)$ denote the Galois group
of $\tilde F=F(\xi)$ over $F$;
by Lemma 6 and its corollary, $|\Gamma|$ is odd.

Let $A^*$ denote the set of all the irreducible characters of $A$ over
$\tilde F$ (i.e. all the homomorphisms $\chi:A\to\tilde F^\times$).
With the usual multiplication of functions, $A^*$ is an abelian group
and $A^*\cong A$. Note that for any integer $k$,
$$
 \chi^k(a)=\chi(a^k)\,,\qquad \forall~ \chi\in A^*\,,~ a\in A\,.
$$
in particular, $\chi^{-1}(a)=\chi(a^{-1})$.

Each $\chi\in A^*$ corresponds exactly one irreducible
module $\tilde Fe_\chi$ of $\tilde FA$, where
$$
 e_\chi=\frac{1}{n}\sum_{a\in A}\chi(a^{-1})a
$$
is a primitive idempotent of the algebra $\tilde FA$.
And we have the direct decomposition of irreducible $\tilde FA$-modules:
$$
 \tilde FA=\bigoplus_{\chi\in A^*}\tilde Fe_\chi\,.
$$
For $\chi,\;\psi\in A^*$ and $\lambda,\;\mu\in\tilde F$,
the standard inner product
$$
 \langle\lambda e_\chi,~\mu e_\psi \rangle
 =n\lambda\mu\cdot(\chi|\psi^{-1})\,,
$$
where $(\chi|\psi^{-1})$ denotes the usual inner product of characters:
$$
 (\chi|\psi^{-1})=\frac{1}{n}\sum_{a\in A}\chi(a)\psi^{-1}(a^{-1})
 =\frac{1}{n}\sum_{a\in A}\chi(a)\psi(a)\,.
$$
By the orthogonal relations of characters,
$$
 \left\langle\tilde F e_\chi,~\tilde F e_\psi\right\rangle
 =\cases{\tilde F, & if $\chi=\psi^{-1}$,\cr 0, & otherwise.}
$$
Any submodule $\tilde C$ of $\tilde FA$ corresponds exactly to a subset
$B\subseteq A^*$ such that
$$
 \tilde C=\bigoplus_{\chi\in B}\tilde Fe_\chi\,.
$$
Thus
$$\tilde C^\bot=\bigoplus_{\psi\notin B^{-1}}\tilde Fe_\psi\, $$
where $B^{-1}:=\{\chi^{-1}\mid\chi\in B \}$;
in particular, $\tilde C$ is self-orthogonal code if and only if
$B\cap B^{-1}=\emptyset$.

Recall that $\Gamma={\rm Gal}(\tilde F/F)$ is a cyclic group
generated by the following automorphism
$$ \gamma:~
 F(\xi)\longrightarrow F(\xi)\,,\quad \lambda\longmapsto\lambda^q\,.
$$
The group $\Gamma$ acts on $\tilde F$
hence acts on the ring $\tilde FA$ in the following way:
$$
 \gamma\Big(\sum_{a\in A}\lambda_a a\Big)=
 \sum_{a\in A}\gamma(\lambda_a) a\,,\qquad \forall~
 \sum_{a\in A}\lambda_a a\in\tilde FA.
$$
We denote $(\tilde FA)^\Gamma$ the subring
consisting of all the $\Gamma$-fixed elements of $\tilde FA$.
It is obvious that
$
 (\tilde FA)^\Gamma =FA\,.
$

And $\Gamma$ acts on the set $\{e_\chi\mid\chi\in A^*\}$
of the primitive idempotents of $\tilde F A$:
$$\gamma(e_\chi)=
 \gamma\Big(\frac{1}{n}\sum_{a\in A}\chi(a^{-1}) a\Big)=
 \frac{1}{n}\sum_{a\in A}\gamma(\chi(a^{-1}))a
 =e_{\gamma(\chi)}\,,
$$
where $\gamma(\chi)\in A^*$ is the composition homomorphism
$$
 A\buildrel\chi\over\longrightarrow \tilde F
  \buildrel\gamma\over\longrightarrow \tilde F\,,\quad
  a\longmapsto\gamma(\chi(a))=(\chi(a))^{q}\,,
$$
i.e. $\gamma(\chi)=\chi^q$.
In this way, $\Gamma$ acts on the abelian group $A^*$.

On the other hand, $H$ acts on the ring $\tilde FA$:
$$
 h\Big(\sum_{a\in A}\lambda_a a\Big)=
 \sum_{a\in A}\lambda_a h(a)\,,\qquad \forall~
 \sum_{a\in A}\lambda_a a\in\tilde FA\,.
$$
Similarly, $H$ acts on
the set $\{e_\chi\mid\chi\in A^*\}$
of the primitive idempotents of $\tilde F A$:
$$ h(e_\chi)=
  h\Big(\frac{1}{n}\sum_{a\in A}\chi(a^{-1}) a\Big)
  =\frac{1}{n}\sum_{a\in A} \chi(a^{-1})h(a)
  =\frac{1}{n}\sum_{b\in A} \chi(h^{-1}(b^{-1}))b=e_{h(\chi)}\,,
$$
where $h(\chi)\in A^*$ is the composition homomorphism
$$
 A\buildrel h^{-1}\over\longrightarrow A
  \buildrel\chi\over\longrightarrow \tilde F\,,\quad
  a\longmapsto\chi(h^{-1}(a))\,.
$$
In this way, $H$ acts on the abelian group $A^*$.

In a word, $\Gamma\times H$ acts on the ring $\tilde FA$, and
the action induces the action of $\Gamma\times H$ on the abelian group $A^*$.

Let $C\le FA$ be an $H$-stable submodule; denote $\tilde C=\tilde F\otimes_F C$.
Then $\tilde C$ is a both $H$-stable and $\Gamma$-stable submodule of $\tilde FA$
such that $\tilde C^\Gamma=C$.
Let $B\subset A^*$ be the subset such that
$\tilde C=\bigoplus_{\chi\in B}\tilde F e_\chi\,.$
Since $\tilde C$ is $H$-stable, we see that $B$ is $H$-stable;
and similarly, $B$ is $\Gamma$-stable.
So $B$ is a $(\Gamma\times H)$-stable subset of $A^*$.

Conversely, if $B$ is a $(\Gamma\times H)$-stable subset of $A^*$,
then $\tilde C=\bigoplus_{\chi\in B}\tilde Fe_\chi$ is a
$(\Gamma\times H)$-stable submodule of $\tilde FA$, and
$\tilde C^\Gamma$ is an $H$-stable submodule of $FA$.

Let $\Omega$ be a non-trivial $(\Gamma\times H)$-orbit of $A^*$,
i.e. $1\notin \Omega$.
Let $\chi\in \Omega$, then $\chi\ne 1$
hence the order of $\chi$ is a power of $p$, say $p^\ell$
(recall that $A^*\cong A$ is an abelian $p$-group).
We claim that $\chi^{-1}\notin \Omega$. Suppose it is not the cases,
then there is $\gamma^i\in\Gamma$ and $h\in H$ such that
$\gamma^i h(\chi)=\chi^{-1}$, and
$$h(\chi)=\gamma^{-i}(\chi^{-1})=\chi^{(-1)(-q^i)}=\chi^{q^i}\,;$$
thus $\langle\gamma\rangle\times\langle h\rangle$
acts on the cyclic group $\langle\chi\rangle$ of order $p^\ell$,
and $\gamma^i h$ acts on $\langle\chi\rangle$ as the
nvolution $\chi\mapsto\chi^{-1}$;
but the automorphism group
${\rm Aut}(\langle\chi\rangle)$ is a cyclic group,
hence the product $\gamma^i h$ of the two automorphisms
$\gamma^i$ and $h$ of odd order still has odd order;
it contradicts to that the $\chi\mapsto\chi^{-1}$ is an involution.

The involution $\tau: A^*\to A^*$, $\chi\mapsto \chi^{-1}$,
commutes with both $\Gamma$ and $H$ clearly.
So $\tau$ permutes all the $(\Gamma\times H)$-orbits of $A^*$.
For any non-trivial orbit $\Omega\ne\{1\}$,
since $\tau(\chi)\notin \Omega$ for any $\chi\in \Omega$,
the subset $\tau(\Omega)$ is an orbit different from $\Omega$.
Thus we can partition all the non-trivial orbits into two
collections $B$ and $B^{-1}=\{\chi^{-1}\mid\chi\in B\}$, and we get
the disjoint union
$$ A^*=\{1\}\bigcup B\bigcup B^{-1}\,. $$
Then the code $\tilde C=\bigoplus_{\chi\in B}\tilde Fe_\chi$
is $H$-stable and $\tilde C^\bot=\tilde C\oplus\tilde F$;
hence the code $C=\tilde C^\Gamma$ of $FA$ is $H$-stable and
$C^\bot=C\oplus F$.

\smallskip\noindent{\bf Theorem 3.}\quad\it
Let $G$ be a finite group of odd order,
and $X$ be a finite transitive $G$-set and $n=|X|$.
Assume that $q=|F|$ is coprime to $n$,
and the order of $q$ in the multiplicative group ${\Bbb Z}_n^\times$
is odd. Then there is a permutation code $C\le FX$ such that
$C^\bot=C\oplus F$.\smallskip\rm

{\bf Proof.}\quad We prove it by induction on the order of $G$.
It is trivial for $|G|=1$. Assume $|G|>1$.
Let $x_1\in X$ and denote $G_1$ the stabilizer of $x_1$ in $G$.
Then $G_1$ is a subgroup and $FX={\rm Ind}_{G_1}^G(F)$.
Since $G$ is solvable by Feit-Thompson Odd Theorem,
a minimal normal subgroup $A$ of $G$ is an
elementary abelian $p$-group, where $p$ is a prime.
Since $A$ is normal, the product $AG_1$ is a subgroup of $G$.
There are three cases.

Case 1: $AG_1=G_1$. Then $A\subseteq G_1$, and hence
$A$ is contained in every conjugate of $G_1$ as $A$ is normal.
Thus $A$ acts trivially on $X$, and $X$ is a $G/A$-set and
$FX$ is a permutation module over $G/A$. Since $|G/A|<|G|$,
the conclusion holds by induction.

Case 2: $AG_1=G$. Since $A\cap G_1$ is both normal in $G_1$ and in $A$,
we have that $A\cap G_1$ is normal in $AG_1=G$;
but $A$ is a minimal normal subgroup of $G$, so $A\cap G_1=1$.
Then we have a bijection
$$
 \beta:~ A\longrightarrow X\,,\quad a\longmapsto a(x_1)\,.
$$
Let $A$ acts on $A$ by left translation,
and let $G_1$ acts on $A$ by conjugation; hence
$G=AG_1$ is mapped into the group ${\rm Sym}(A)$
of all the permutations of $A$:
$$
 (bh)(a)=bhah^{-1}\,,\qquad\forall~a,b\in A,~h\in H\,.
$$
Noting that $G_1$ stabilizes $x_1$, we have that
$$
 \beta\Big((bh)(a)\Big)=(bhah^{-1})(x_1)=bha(x_1)=(bh)\beta(a)\,.
$$
Thus, mapping $bh\in G$ to the permutation $a\mapsto bhah^{-1}$ of $A$
is an action of $G$ on $A$, and $\beta$ is an isomorphism of $G$-sets.
Then $n=|A|$ hence $p|n$, so $p$ is coprime to $q$,
and by the assumption of the lemma,
the order of $q$ in ${\Bbb Z}_p^\times$ is odd
(see Lemma 6). The conclusion is derived from Lemma 8.

Case 3: $G_1\lneqq AG_1\lneqq  G$. In this case,
$$
FX\cong {\rm Ind}_{G_1}^G(F)={\rm Ind}_{AG_1}^G{\rm Ind}_{G_1}^{AG_1}(F)\,.
$$
Let $Y=\{gx_1\mid g\in AG_1\}$, which is an $AG_1$-set and
${\rm Ind}_{G_1}^{AG_1}(F)\cong FY$.
By induction, there is a code $D\le FY$ such that $D^\bot=D\oplus Fe_Y$
where $e_Y=\sum_{y\in Y}y$.
Turn to the permutation module $FX={\rm Ind}_{AG_1}^G(FY)$,
by Lemma 7, we have
$$
 {\rm Ind}_{AG_1}^G(D)^\bot={\rm Ind}_{AG_1}^G(D^\bot)
 ={\rm Ind}_{AG_1}^G(D\oplus Fe_Y)
 ={\rm Ind}_{AG_1}^G(D)\oplus{\rm Ind}_{AG_1}^G(Fe_Y)\,.
$$
Noting that, $Fe_Y$ is the unique trivial module of $FY$, and
$${\rm Ind}_{AG_1}^G(Fe_Y)=\bigoplus_{t\in G/AG_1}t\otimes Fe_Y\,;$$
by induction again, there is a code
$E\le {\rm Ind}_{AG_1}^G(Fe_Y)$ such that
$$ {\rm Ann}_{{\rm Ind}_{AG_1}^G(Fe_Y)}(E)=E\oplus Fe_X\,, $$
where $e_X=\sum_{x\in X}x$.
So we can write ${\rm Ind}_{AG_1}^G(Fe_Y)=E'\oplus E\oplus Fe_X$,
and have
$$
 {\rm Ind}_{AG_1}^G(D)^\bot
 ={\rm Ind}_{AG_1}^G(D)\oplus{\rm Ind}_{AG_1}^G(Fe_Y)
 ={\rm Ind}_{AG_1}^G(D)\oplus E'\oplus E\oplus Fe_X\,.
$$
Let
$$ C={\rm Ind}_{AG_1}^G(D)\oplus E $$
which is a permutation code of $FX$ and
\begin{eqnarray*}
 C^\bot&=&{\rm Ind}_{AG_1}^G(D)^\bot\bigcap E^\bot
 ={\rm Ann}_{FX}\Big({\rm Ind}_{AG_1}^G(D)\Big)\bigcap
   {\rm Ann}_{FX}(E)\\
 &=&\Big({\rm Ind}_{AG_1}^G(D)\oplus E'\oplus E\oplus Fe_X\Big)
   \bigcap{\rm Ann}_{{\rm Ind}_{AG_1}^G(D)\oplus E'\oplus E\oplus Fe_X}(E)\\
 &=&\Big({\rm Ind}_{AG_1}^G(D)\oplus E'\oplus E\oplus Fe_X\Big)
   \bigcap\Big({\rm Ind}_{AG_1}^G(D)\oplus E\oplus Fe_X\Big)\\
 &=& {\rm Ind}_{AG_1}^G(D)\oplus E\oplus Fe_X\\
 &=& C\oplus Fe_X\,.
\end{eqnarray*}

\smallskip
As a consequence of Theorem and Lemma 5 (cf. its remark),
we get the followings at once.

\smallskip\noindent{\bf Corollary 2.}\quad\it
Assume that $q=|F|$ is even and $|G|$ is odd
and $X$ is a transitive $G$-set and $n=|X|$.
If the order of $q$ in the multiplicity group ${\Bbb Z}_n^\times$ is odd,
then there is a self-dual extended code of $FX$.
\smallskip\rm

\noindent{\bf Corollary 3.}\quad\it
Assume that $|G|$ is odd and $X$ is a transitive $G$-set and $n=|X|$.
If $q=|F|$ is coprime to $n$ and
the order of $q$ in the multiplicity group ${\Bbb Z}_n^\times$ is odd,
and $-n$ has square root in $F$,
then there is a self-dual extended code of $FX$.
\smallskip\rm

\end{document}